\UseRawInputEncoding
\documentclass[12pt,letterpaper]{article}
\usepackage{amsmath,amssymb,pgf,pgfarrows,pgfnodes,float,appendix, hyperref}
\usepackage{graphicx}
\usepackage{subfigure}
\usepackage[margin=0.9in]{geometry}

\newcommand{\be}{\begin{equation}}
\newcommand{\ee}{\end{equation}}
\newcommand{\bea}{\begin{eqnarray}}
\newcommand{\eea}{\end{eqnarray}}

\title{{\rm\footnotesize \qquad \qquad \qquad \qquad \qquad \ \qquad \qquad \qquad \ \ \ \ \ \                 RUNHETC-2026-23}\vskip.5in  On the Unitarity of the Gravitational S-matrix in High Dimension}
\author{Tom Banks\\
Department of Physics and NHETC\\
Rutgers University, Piscataway, NJ 08854\\
E-mail: \href{mailto:tibanks@ucsc.edu}{tibanks@ucsc.edu}
\\
\\
\\
\\
}
\date{}
\begin{document}
\maketitle

\begin{abstract} We argue that for finite energy windows, the final states in gravitational scattering in dimension $d > 4$ are normalizable coherent states in Fock space.  However, as the center of the energy window goes to infinity, black hole physics predicts that these states become orthogonal to every state with a finite number of particles.  Given that the spectral measure in energy is determined by Poincare invariance, the S-matrix cannot be a unitary operator in Fock space, despite having finite matrix elements in Fock space, and satisfying perturbative unitarity, to all orders in string perturbation theory. We identify regimes in the BFSS matrix model\cite{bfss} and the definition of the S-matrix as the limit of CFT correlators\cite{polchsuss}, which point to the same conclusion.  We review a scattering theory based on the quantum mechanics of a finite number of fermionic oscillators, whose algebra formally converges to the Super-Poincare covariant Awada-Gibbons-Shaw\cite{ags} algebra, and argue that a certain class of limiting states on that algebra satisfy all the properties required by physical unitarity in the algebraic formulation of quantum mechanics.  The only missing ingredient for a consistent theory is a proof that the S matrix amplitudes themselves are Poincare invariant.   We provide suggestive arguments, but no real proof, that this is so.  
 \end{abstract}
\maketitle

\section{Infrared Properties of High Dimensional Gravitational Scattering}

It is well known that the perturbative gravitational S-matrix in $4$ space-time dimensions is infrared divergent\cite{weinberg}.  Recently, even the Fadeev-Kulish\cite{FK} construction of IR finite amplitudes has been called into question\cite{waldetal}.  In higher dimensions the IR problems seem to go away, and perturbative superstring theory seems to provide us with amplitudes that satisfy perturbative unitarity in Fock space to all orders in perturbation theory.  However, the fact that the string perturbation series is not Borel summable\cite{shenker} means that this cannot be turned into a proof that there is a definition of a non-perturbative completion of the series that obeys Fock space unitarity exactly.

The purpose of the present note is to establish several different goals.  We will first review and sharpen arguments that all known perturbative and non-perturbative effects in gravitational theories at finite energy lead, in space-time dimension $d \geq 5$ to final states that are at worst normalizable coherent states of gravitons in Fock space.   However, we'll also argue that if we restrict the scattering operator $S(E)$ to a microcanonical window $ E_0 \pm \Delta$ with some small width, then as $E_0 \rightarrow\infty$ many of the final states in scattering initiated by states with a fixed number of particles, have an order one probability to have zero overlap with any state with a finite number of particles. Since the measure in $E$ is fixed by symmetry, this means that the non-perturbative S matrix cannot be unitary in Fock space.

In the next section of the paper we explain how the phenomenon motivated by black hole physics and soft radiation theorems in Section 2, can arise mathematically in non-perturbative definitions of the S matrix defined by\cite{bfss} and \cite{polchsuss}.  

Finally, we'll review a finite dimensional quantum mechanics problem, which has a large N limit with an asymptotically conserved energy variable.  The scattering problem is defined by constraints on the irreducible representation space of a finite number of canonical fermionic oscillators, and a sequence of time dependent Hamiltonians built from those oscillators.  The algebra of operators formally converges to the local supersymmetry algebra of Awada Gibbons and Shaw\cite{ags}, defined on a null momentum cone dual to Penrose's null infinity.  The generators are half-measures on the cone, and the finite dimensional constraints translate to statements about were the half-measures supported at $P  = 0$ vanish.   The limiting algebra obviously exists in the usual algebraic QFT sense (with half-measureable spinors instead of smooth functions as test functions ), but it is unclear whether the constraints are realizable in a representation on a separable Hilbert space.  We argue that physical unitarity, the appropriate (fuzzy) generalization of a local normal state on the limiting algebra, exists, in which the S matrix is unitary.  The challenge that remains is to prove that it is Poincare invariant.  

\section{Infrared Behavior in $d > 4$}

Consider two body gravitational scattering at center of mass energy $E$ and impact parameter $b$ in dimension $ d > 4$ .  We can ask, when $b > R_s (E)$ the Schwarzschild radius of the c.m. energy, what the amount of energy is that goes into ``soft" gravitational radiation, where soft is defined by saying that the deviation from the outgoing trajectories of the final particles is negligible.  The answer is given by Weinberg's soft theorem\cite{weinberg} and is well known to be
\begin{equation} E_{IR} \sim E  (\frac{R_s (E)}{b})^{d - 3}   . \end{equation}
This is very different than the $d = 4$ case, where the power of $b$ is anomalous and there's an extra logarithmic enhancement.  Note that in all cases, when expressed in terms of Newton's constant, this is quadratic in $E$.  Weinberg's soft theorem implies that the soft graviton emission amplitudes exponentiate and give the same result as a classical coherent state of gravitons.  For $d \geq 5$ that state is normalizable in Fock space.  

However, it's also consequential that as $E/M_P \rightarrow \infty $ with $b$ any fixed multiple of $R_s (E)$ , $E_{IR}$ goes to infinity.  Phase space considerations tell us that the typical soft graviton energy is $\sim 1/b$.  So the average number of soft gravitons in the state must be growing like $E b > E R_s (E)$.  Indeed we find that
\begin{equation} \langle N \rangle \sim (\frac{R_s (E) /L_P})^{d-2} (\frac{b}{R_s (E)})^{7 - 2d} . \end{equation}
The final states in high energy two body scattering, in a growing range of impact parameters are becoming more and more orthogonal to the finite particle number region of Fock space.   Similar considerations are valid for analogous kinematic regimes of high energy multiparticle scattering, where the particles remain fairly far apart in units of the Schwarzschild radii of their sub-energies.   

Things get even more strange if we contemplate the possibility\cite{penroseetal} of black hole creation in high energy particle scattering, which is by now considered part of standard lore.  Starting from a state of a few high energy particles, with several sub-energies much larger than the Planck scale, we can create several large black holes, which orbit around each other, emit gravitational radiation, perhaps coalesce, and in the classical approximation leave over a few huge isolated black holes.   Semi-classical dynamics predicts that these will decay by Hawking radiation.  Soft theorems again predict that the graviton emission in all of these complicated processes correspond to normalizable states in Fock space.  Hawking radiation predicts a thermal density matrix, which is again a normalizable, though impure, Fock space state.  It is now standard lore that the final state is actually pure, but will have all of the coarse grained statistical properties of the thermal density matrix.  In particular, since the Hawking temperature of these large black holes tends to zero, these states become orthogonal to all finite particle number states in Fock space in the infinite energy limit.  The S matrices in high energy windows cease to be unitary operators in Fock space.  

To make these remarks more quantitative, note that soft theorems predict that for impact parameter $b$ larger than but commensurate with the Schwarzschild radius $R_s (E)$ there is, according to the soft theorem, an $o(1)$ probability for the final state to be well approximated by a coherent state of a large number of gravitons of wavelength $ \sim b$.  The sum of the squares of the overlaps of this state with Fock states containing $\leq M$ particles is
\begin{equation} \Sigma_{n\leq M} | \langle n | coherent (E) \rangle|^2 \sim e^{- c(L,M) (\frac{E}{M_P})^{\frac{d-2}{d-3}}} , \end{equation} where $b = L R_s (E)$ and $c(L,M)$ is an order one constant.  This calculation is done in semi-classical gravity, in a regime that is completely safe from corrections due to its ultraviolet completion.  Despite the high energy, all wavelengths are long and no black holes are formed.   We note that even if $M $ is allowed to grow like $N(E)^a$ with $a < 1$ , where $N(E)$ is the mean occupation number in the coherent state, then the contribution is still exponentially suppressed.  

We can get a similar estimate if we accept the once heretical suggestion\cite{penroseetal} that black holes are formed with finite probability in high energy scattering with impact parameter $ b < R_S (E)$.  We combine this with the assumption that the typical pure state produced by the Hawking decay of a black hole has the same statistical properties as Hawking's thermal ensemble.  Thus a finite probability piece of the few particle high energy scattering amplitude has the same over lap with the part of Fock space with $\leq M$ particles as a thermal state at the Hawking temperature.  This gives
\begin{equation} \Sigma_{n\leq M} | \langle n | BH (E) \rangle|^2 \sim e^{- c_{BH} (M) (\frac{E}{M_P})^{\frac{d-2}{d-3}}} , \end{equation} where again the $M$ dependence is weak.  Thermal fluctuations actually have less support at small particle number than coherent state fluctuations, but the essential message is the same.  There is a high energy/IR connection that makes the Fock space picture of the Hilbert space incomplete even in high dimension.  It follows from mild assumptions about regimes where a UV complete theory of quantum gravity should agree with calculations we know how to do with semi-classical methods\footnote{In a UV complete model like perturbative superstring theory, there will be all sorts of other kinematic regimes, dominated by string production, D-brane production, Gross-Mende re-summation, {\it etc.}. The eikonal and black hole production regimes we discuss are universal to all models of quantum gravity and their coarse grained properties are independent of the details of UV completion.}.  In the next section we'll explore where these regimes arise in non-perturbative models of flat space gravitation.  It's worth reminding the reader again that perturbative unitarity of UV complete string theory amplitudes does not say anything, either positive or negative, about these arguments.  For Borel summable perturbation series, perturbative unitarity is enough to prove that the Borel sum is a unitary operator, but the string perturbation series is not Borel summable.  Moreover string perturbation theory clearly re-sums to the Weinberg soft theorem for $L > 1$ and manifestly breaks down when $ b \sim R_s (E)$, so it does not contradict anything we have said.  

Let's emphasize that what we have here is a proof by contradiction.  We assume a Poincare invariant S-matrix for gravitons in Fock space.  Unitarity and Poincare invariance imply that the S-operator integrated over any small energy window with a known measure is a unitary operator.  Weinberg's soft theorem allows us to evaluate certain matrix elements of that operator as products of an eikonal phase and production of a soft graviton coherent state.  Despite the fact that $E \gg M_P$ we're in a regime where corrections to this effective field theory calculation are expected to be negligible.  As the center of the energy window is taken to infinity, the overlap of this final state with all states of finite particle number in Fock space goes to zero exponentially.  A similar conclusion is reached for smaller impact parameters if we assume that black holes are formed and that the final states of their decay have the statistical properties of Hawking's thermal density matrices.  These results are compatible with perturbative unitarity in Fock space in an ultraviolet complete model of quantum gravity, as long as the perturbation series in NOT Borel summable\footnote{The closest one has to a proof of unitarity for Borel summable perturbation series is in\cite{simons}.}.  These arguments do not tell us that the S matrix is not a unitary operator, only that it is not unitary in Fock Space.  In a later section we'll argue that there is an algebraic scattering theory which incorporates the physical constraints of unitarity.  The question of whether there is a unitary operator for gravitational scattering in some separable Hilbert space will not be addressed. 

It's instructive to compare these considerations to similar calculations for Yang-Mills theory in higher dimensions.  Here too we find that soft theorems predict large amounts of energy goes into particles with wavelengths of order the impact parameter as the energy grows.  However, in this case we run up against the UV cut-off of the theory.  A case that's fairly easy to understand is one in which we obtain $5$ dimensional Yang-Mills theory by compactifying the $(2,0)$ superconformal field theory on a circle.  Here, when the energy reaches the compactification scale we begin to excite Kaluza-Klein modes.  This will happen below some impact parameter of order the compactification radius of the circle in the fifth spatial dimension. 

Everything takes place in the Hilbert space of the compactified CFT.  The full set of asymptotic states includes particles with Kaluza-Klein charge, which are 4 dimensional Yang-Mills instantons.  The full S matrix is not unitary on the original Fock space of gluons but there is no argument that unitarity is violated on the larger Fock space.  It's clear that in this case the exact presentation of the Hilbert space is the familiar one in terms of operators acting on the CFT ground state.  The high energy spectrum of the theory is entirely dominated by conformally invariant six dimensional physics and does not have an S-matrix interpretation in five or six dimensions.  It's clear that if we compactify four dimensional space on a four sphere (something we can't do for gravity) , then the only proper description of the energy spectrum, particularly at high energies, is in terms of the standard CFT description of operators acting on a state that preserves a maximal subgroup of the conformal group.  Extracting a unitary S matrix on the Fock space of Yang-Mills bosons and instantons from this rigorous description is, as far as the present author knows, an unsolved mathematical problem.  

One might also worry about similar problems for ordinary renormalizable theories with Goldstone bosons. Indeed, the algebraic scattering theory that we will mimic in the final section of this paper was developed to study precisely that situation.  A more pedestrian solution of course is to add a small symmetry breaking perturbation, define the S-matrix and take limits of amplitudes as the perturbation goes to zero.  There's no analog of this procedure for gravity.

\section{AdS/CFT and BFSS Matrix Models of Quantum Gravity}

AdS/CFT and the BFSS Matrix models are non-perturbative quantum mechanical models, which each claim to be able to compute the Minkowski space S-matrix by taking limits of a subset of amplitudes in the original model.  In both cases these claims have been verified for a subset of Fock space amplitudes.  The issue of soft gravitons has been addressed in a number of papers for the BFSS model\cite{softbfss}, as we will review below.  For AdS/CFT it was discussed, to my knowledge, only in\cite{tbwfads2} and remarks referring to that paper in later work of the present author.  The essence of the argument is as follows:  the unitarity equations for connected correlation functions of a conformal field theory on the sphere\cite{nishijima} involve connected $n-$point functions for all $n$.  When one takes the limit $R_{AdS}/L_P\rightarrow\infty$ in the prescription of\cite{polchsuss}, with some finite number of boundary operators focussed on the ``arena" causal diamond that becomes the Penrose diagram of Minkowski space in the limit, there are an infinite number of connected correlators between these operators and others which are NOT focussed on the arena.  In the limit, in the Witten diagram approximation, the dominant contributions to these correlators come from gravitons and other massless particles exchanged between the operators localized in the arena, and the propagators of other particles outside the arena.  It's clear that some subset of these correlators give rise to the soft graviton contributions to the Minkowski S-matrix.

An alternative argument, which leads to the same conclusion, follows if one accepts the premise of\cite{tbwfads2} that one can view the shells of a tensor network approximation to a CFT as the bifurcation surfaces of finite area causal diamonds along a particular time-like geodesic in AdS space.  The Polchinski-Susskind arena diamond is then identified with the domain of dependence of the central node of the network.  It's clear from this picture that in the vacuum state of the CFT, the entanglement entropy of this diamond goes to infinity as the AdS radius is taken to infinity in Planck units.  Thus, the entropy of the Minkowski vacuum has a soft graviton divergence even in high dimensions.  

A much more detailed picture of how soft gravitons arise has emerged from work on the BFSS model\cite{softbfss}.  We can only give a very abbreviated summary of the arguments here.  Two gravitons with relative transverse velocity $\vec{v}$ are represented by matrix blocks of size $N_{1,2}$ which are finite fractions of the total matrix size $N$.  The matrix model energy of the incoming state is of order $1/N$.  Now consider any one of the $\sim N_1 N_2$ off diagonal coordinates at impact parameter $b$.   Semi-classically, this is a time dependent harmonic oscillator with frequency
\begin{equation} \omega \sim \sqrt{b^2 + v^2 t^2} . \end{equation}  The adiabatic probability to excite an off diagonal mode is
\begin{equation} \frac{\dot{\omega}}{\omega^2} \sim \frac{v}{b^2} . \end{equation}  The transverse velocity scales as $1/N$ as $N \rightarrow\infty$ , while the impact parameter is kept fixed. So the number of excited off diagonal modes scales like $N$.  There isn't enough energy in the initial state to actually support real production of these modes, and of course asymptotically the transverse separation goes to infinity and it becomes harder and harder to excite them.  Instead, the energy must be dissipated by splitting off many little  blocks with small transverse momenta so that the total matrix model energy is of order $1/N$.  These are the soft gravitons.  Details of these arguments can be found in\cite{softbfss}.  

Another explicit exercise that one can do in the BFSS model is to compute the limiting partition function at temperatures $N\beta$.  This will include the usual contribution from supergraviton Fock space, which comes from configurations on the moduli space with block diagonal matrices whose block size is a finite fraction of $N$, and whose transverse momentum is finite.  But there are also contributions from sectors with block sizes scaling like $N^a$ with $a < 1$ and $|p_{\perp}|^2 \sim N^{-1 + a}$ for all $ a > 0$. This tells us nothing about the dynamical relevance of these states, but it does say that the Wilsonian argument that separates the $o(1/N)$ energies from matrix model junk that is not relevant to M theory, is not sufficient to eliminate these states.

\section{An Algebraic Quantum Scattering Theory} 

For simplicity, we will restrict attention to $d = 11$.  String theory and AdS/CFT have taught us that models of quantum gravity in asymptotically flat space-time must be exactly supersymmetric.  While this is not (yet) a mathematical theorem, there is an overwhelming amount of empirical evidence in favor of this conjecture.  For $d < 11$ there are multiple models, and many of them have a spectrum of stable massive BPS particles, and sometimes stable non-BPS states carrying discrete ``K-theory" like symmetry charges.  A discussion for general $d$ would have to take all of these complications into account.  In $11$ dimensions, we believe there is a unique consistent theory, containing only a single supermultiplet of massless particles, the supergraviton. 
Many years ago, Awada Gibbons and Shaw\cite{ags} derived an asymptotic algebra from supergravity in any number of space-time dimensions.  Their derivation had nothing to do with soft theorems or memory effects, but merely reflected the fact that the covariant conservation law of the supercurrent becomes an ordinary conservation law at null infinity, because of the asymptotic boundary conditions on the fields.  This results in fermionic operators describing the flow of supercurrent through the boundaries of the Penrose diamond.  In the papers\cite{tbsuperBMS} we found a translation of the AGS algebra to the null momentum cone $P^2 = 0$.  It is
\begin{equation} [Q^{\pm}_a (P) , Q^{\pm}_b (P^{\prime})]_+ = (\gamma^m )_{ab}  P_m \delta (P\cdot P^{\prime}) . \end{equation} 
\begin{equation}  P_m (\gamma^m )_{ab} Q^{\pm}_a (P) = 0 . \end{equation} 
The $\pm$ superscripts refer to fields with support for $ \frac{P^0}{|P^0|} = \mp 1$, which we associate with the outgoing and ingoing massless states, respectively.  The fields at $P = 0$ are shared between the two algebras.  It's clear from the algebra that we should think of these fields as operator valued half-measures (half-densities) on the null cone.  In a moment, we'll introduce a regularized version of this algebra, appropriate to finite causal diamonds.  The regularized version is defined on a finite dimensional Hilbert space and the scattering theory defined on it is defined by considering constraints on the states in that space.  The finite dimensional Hilbert space represents a very large but finite size causal diamond, which is covered by a time symmetric nesting of causal diamonds as in Figure 1.  

\begin{figure}[h]
\begin{center}
\includegraphics[scale=0.5]{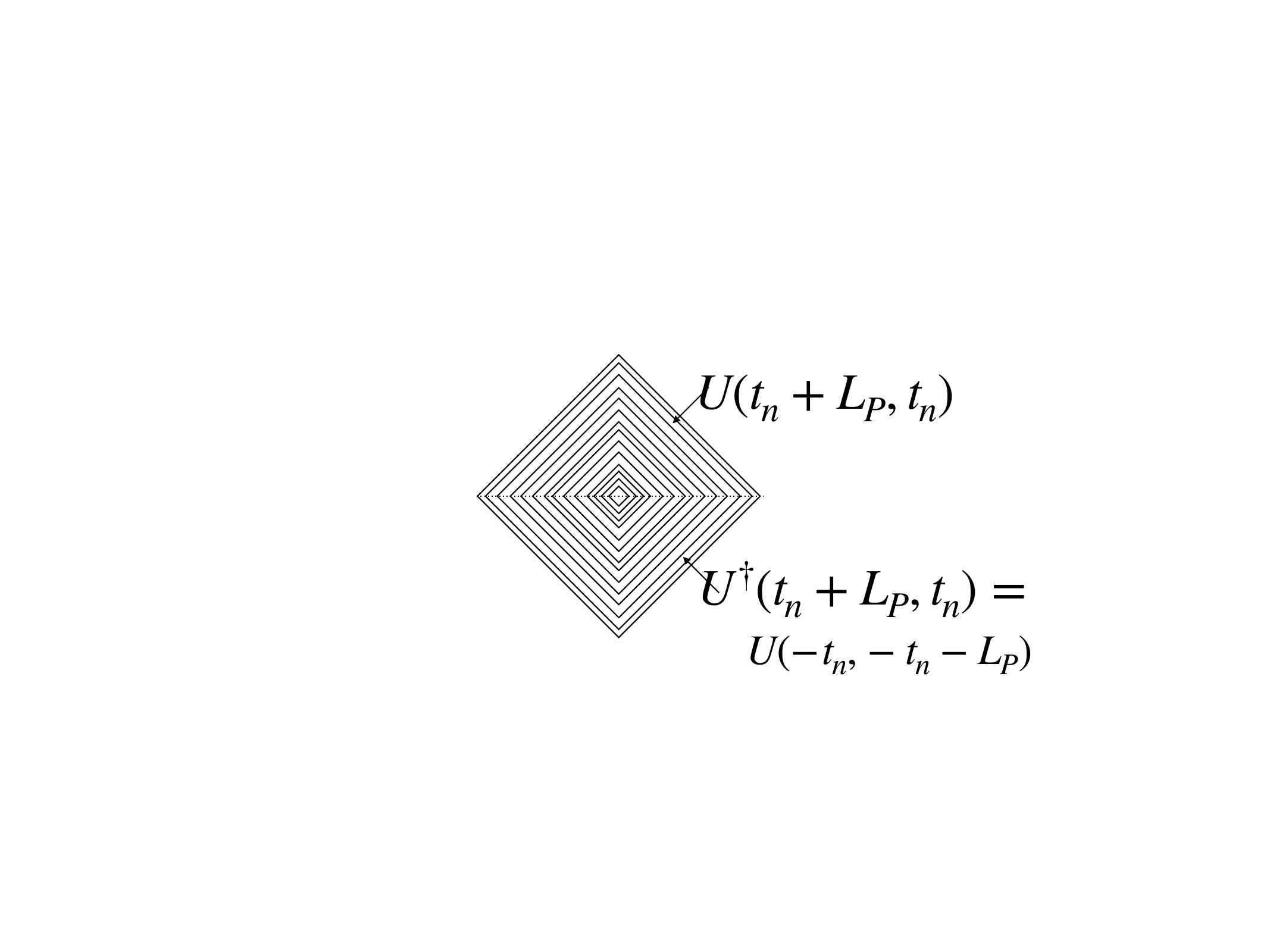}
\caption{Symmetrically Nested Causal Diamonds}
\label{z}
\end{center}
\end{figure}

These represent the diamonds corresponding to proper time intervals $[- t_n, t_n]$ along some geodesic in Minkowski space, with $t_{n+1} = t_n + L_P$.  Each diamond will have a prescribed density matrix $e^{- K_{\diamond}}$ in the {\it empty diamond state} of the system, the analog of the QFT vacuum.  This global state is not pure.  

Scattering states in the regulated system are defined by constraining the q-bits in a manner specified below.  The number of constrained q-bits on the maximal diamond is of order $T/L_P$, the linear size of that diamond and goes to infinity as we approach the Penrose diamond.  We will argue that the dynamics is such that this number becomes an asymptotically conserved quantity, which we identify as energy in the rest frame of the particular geodesic defining the nesting.  

This definition depends on a choice of geodesic and it defines a Hilbert bundle of quantum systems over the space of time-like geodesics on Minkowski space\cite{hilbertbundles}.  The (flat) connection on this bundle is the requirement that whenever two diamonds along different geodesics have a non-trivial intersection, then the maximal diamond in the intersection has a density matrix with the same entanglement spectrum, no matter which quantum mechanical system we use to compute it.  We enforce this condition by insisting that at any given discrete proper time $t$ the time evolution operator along any geodesic factorizes 
\begin{equation}  U(t, t - L_P) = U_{in} (t, t - L_P )\otimes U_{out} (t, t - L_P) . \end{equation}   $U_{in}$ is a unitary embedding of the Hilbert space corresponding to the interval $[t - L_P , 0]$ into that of the interval $[t,0]$, while $U_{out}$ acts on its tensor complement in the full Hilbert space of the interval $[T,-T]$.   We have a similar factorization for the interval $[0, - t]$.  We implement CRT invariance by insisting that the evolution operators for reflected time intervals are Hermitian conjugates of each other.  Note that this implies that the evolution operators $U_{in} (-t - t - L_P)$ are unitary {\it dis-entanglers}, mapping a larger Hilbert space into a smaller one.  Readers familiar with the Tensor Network Renormalization Group will note the similarity between the flow of time here and the flow of scale there, which was exploited in\cite{tbwfads2}.  

$U_{in}$ is required to map the empty diamond density matrix of one diamond into that of the next larger one.  We will see how to do that below.  $U_{out}$ is free to be chosen to make the consistency requirements on overlaps work.  For example, choose at a fixed time $t_0$ along a lattice of synchronized geodesics that are all at relative rest, and such that the maximal area surfaces of adjacent diamonds just touch, the time evolution operators to be tensor products of the same set of unitary inclusions.  There are no overlaps between any of the diamonds up to this point in proper time.  Let this time $t_0$ be the coarse graining scale.  We will defer for the moment the question of whether there is any way, mathematically or ``experimentally" to probe physics at scales shorter than this.  

Up to time $t_0$ we can satisfy the connection condition for our Hilbert bundle, {\it on a lattice of geodesics at relative rest and causally disconnected up until time $t_0$} by taking the full evolution operator to be the tensor product of the evolution operators on individual geodesics on the lattice.  Now let's go to a larger time $ t > t_0$.  The causal diamonds of some pairs of geodesics in the lattice will overlap and the overlap region will contain a causal diamond of maximal area.  We should identify this as an area on the maximal area surface of each individual geodesic.  If the dynamics along each geodesic is invariant under volume preserving maps, and is identical along each geodesic, then the reduced density matrix associated with this subvolume will be the same, regardless of which fiber of the Hilbert bundle we choose to do dynamics in.  Below, we will propose a density matrix for each causal diamond in the {\it empty diamond state}, the analog of the Minkowski vacuum state in QFT, which is invariant under a ``fuzzy" version of volume preserving maps, and a dynamics inside diamonds which preserves this state.  We will also define localized excitations inside diamonds as constrained states characterized by a variable that just counts the number of frozen q-bits relative to the empty diamond state.   Using the relation between entropy and area, this also translates into a statement that, along any given geodesic, is invariant under volume preserving maps.  The geometric information about {\it where} a particular set of frozen q-bits enters and/or exits the boundaries of a given causal diamond is contained only in the consistency conditions between different geodesics, and their positions in the background geometry.  A certain set of frozen q-bits can enter the past boundary of a given diamond at a certain time, only if it exits the future boundary of an adjacent diamond at that time.  The background geometry sets the framework in which all of these compatible quantum systems evolve.  

Thus, at least in a coarse grained way, we can define $U_{out} $ on each geodesic at each time $t > t_0$ to be the tensor product of the $U_{in}$'s for a lattice of non-overlapping diamonds whose maximal surfaces just touch.  The consistency conditions then require that we identify a sub-algebra of the operator algebra of each diamond with the same density matrix according to the dynamics along each geodesic.  We'll see how this can work in an explicit model below.  The consistency conditions on constrained states depend on how the different geodesics are located in the background geometry and they therefore define a trajectory for the constrained variables through the background space-time.   We will argue that our particular choice of variables defines, for large diamonds, a fluctuating Riemannian geometry on the maximal area surface of each diamond, in which the constrained variables have the topology of an annular region on a sphere, surrounding a spherical cap.  The sphere is the background geometry around which the small fluctuations fluctuate.  The fact that these regions always have volume that is a small fraction of the total, going to zero as the size of the diamond goes to infinity, implies that we can always use the invariance group of the dynamics to make these constraint regions symmetric under $SO(9)$.  

Now, going back to the time $t_0$, we examine the Lorentz boost of the frame in which all of the disjoint diamonds are at rest.  Without loss of generality, we can do the boost around the origin.  The boosted diamond at the origin then has overlaps with diamonds in the unboosted lattice at roughly the points
\begin{equation} n_z \in [ - (1+ \cosh \eta)/2, ( 1 + \cosh \eta)/2 ], \ \ n_i \in [-1, 0 , 1]   .\end{equation}   $\eta$ is the rapidity of the boost, and the two different behaviors refer to directions parallel to and orthogonal to the boost direction, with the two signs for the parallel case referring to diamonds in the boost direction or its reflection.  The maximal sizes of the overlap diamonds are given roughly by the formulae
\begin{equation} a_{max} = a e^{-\eta} (1 - \frac{n_z}{\cosh\eta} - |\vec{n_{\perp}}|)    . \end{equation}   Similar but more complicated formulas can be written for the overlaps between the time evolution of diamonds in the two lattices, each with the appropriate Lorentz transformed time variable.

Thus, we argue again that if our prescription for the empty diamond density matrices of diamonds is invariant under volume preserving maps of the maximal volume surface on diamond boundaries, and our description of localized excitations is in terms of a count of frozen q-bits, relative to the empty diamond distribution, then as long as the dynamics inside each diamond is consistent with mapping these density matrices into each other, the conditions on overlaps of diamonds will be satisfied just by matching the labeling of frozen q-bits to the geometrical positions of diamonds in space-time.  This can be summarized by saying that the trajectories of localized objects are determined by following the consistent freezing and unfreezing of q-bits in a set of diamonds that fills space-time.  

Now let us turn to the construction of a system that implements these ideas mathematically, and asymptotes to the AGS algebra at null infinity. 
The constraint $  P_m (\gamma^m )_{ab} Q^{\pm}_a (P) = 0 $ tells us that the 11 dimensional spinors are determined by spinor fields on the sphere $S^9$ , transforming in the $16$ dimensional representation of $SO(10)$.

The formalism outlined in\cite{hilbertbundles} combines the insights of\cite{ted95}\cite{carlip}\cite{solo}\cite{BZ} with the program of\cite{hst} initiated many years ago by 
Fischler and the present author.  The basic idea is to use hydrodynamic information about quantum gravity (QG) gleaned from the Einstein-Hilbert action to conclude that the effective theory on a causal diamond boundary is a cut-off $1 + 1$ dimensional CFT with a central charge proportional to $\frac{A_{\diamond}}{4G_N}$.  The connection to the work of\cite{hst} suggests that this CFT is a theory of free  $1 + 1$ dimensional massles Dirac fermions, in one to one correspondence with eigenspinors of the $9$ dimensional Dirac operator on the sphere $S^9$ up to some eigenvalue cutoff.  The work of Connes\cite{connes} and\cite{tbjk} show that this provides a ``fuzzified" description of the spherical geometry.  We can view the fluctuating $1 + 1$ dimensional fields as describing small fluctuations of the geometry of the maximal volume sphere on the diamond's boundary.  In order to incorporate fast scrambling of perturbations on the horizon\cite{lshpss} and the expected spectral statistics of meta-stable black hole eigenstates, we add a small one parameter $U(N)$ invariant Thirring coupling to the model for the modular Hamilton of the diamond 
\begin{equation} K_{\diamond} = L_0 (\diamond) + g \int {\rm Tr} {J_m J^m} . \end{equation}  The sign of $g$ is chosen so that the non-abelian part of the coupling is marginally irrelevant, and its magnitude scales with the central charge so that the main role of this term is to break the degeneracies in the free fermion spectrum by small amounts\footnote{In\cite{hilbertbundles} a different proposal was presented for enforcing fast scrambling. That proposal was criticized by S. Shenker, because it led to a diamond density matrix with integrable spectrum, contrary to the expectation for black holes.}.  The $U(N)$ invariance becomes invariance under volume preserving maps in the formal limit in which the fuzzy geometry becomes a continuous measure space.  

The time evolution operator connecting two nested diamonds with Planck scale separation between their tips is the unitary embedding
\begin{equation} U_{in} (t, t - L_P , - t + L_P , - t) = e^{i K L_P  (t - L_P)/N_t  } e^{- i KL_P (t)/N_t} . \end{equation} $N_t$ is proportional to $(\frac{A_{\diamond}}{4G_N})^{\frac{1}{9}}$ and will be defined more precisely below. In specifying time evolution, we are referring to a particular time coordinate, which is the proper time along the geodesic defining the nesting of diamonds.  The scaling of the Hamiltonian in the formula above is such that the natural time scale for evolution is of order $N_t L_P $, the entire proper time in the diamond. Note however that the Hamiltonian is time dependent, and that as we go in from $- N_t L_P$ many degrees of freedom do not evolve under the time dependent Hamiltonian at all.  The time scale for evolution of those that do remain in the smaller diamonds speeds up.  

As noted above, we do not yet have a full mathematical solution for $U_{out}$ obeying the consistency condition for the Hilbert bundle connection.  However, the proposal above obeys the intuitive rules that we proposed for implementing that condition.  The different fermion fields in each diamond are equivalent to each other and the constraints defining localized excitations entering a diamond (see below) just depend on the number of those fields that we freeze.  We make the assumption that the initial conditions in the empty diamond state have prepared the density matrix in the empty diamond density matrix, so that this time evolution is consistent in that state.  

Now comes the crucial part of our discussion, the definition of scattering states.  The counting of eigenspinors of the $9$ dimensional Dirac operator up to some cutoff is the same as that of totally anti-symmetric tensors of rank $9$, with indices that run from $1$ to some integer $n \propto R_{\diamond}/L_P$.  We denote the maximal diamond size by $N$, so that the Minkowski limit is $N \rightarrow \infty$. 

To define localized excitations, we first note that energy and momentum are not connected to the time dependent discretized Hamiltonian evolution we've outlined above.  Rather, they'll be emergent conservation laws, sharply defined only in the $N \rightarrow\infty$ limit.  This coincides with the intuition from classical GR that they are associated with asymptotic symmetries.  The covariant conservation law of the ``matter" stress tensor, becomes an asymptotic conservation law of a stress tensor including gravitons, and the energy momentum vector components are the charges associated with these conserved currents at null infinity.  In this sense, despite the usual wisdom about the Bondi-Metzner-Sachs algebra in higher dimensions, there are charges localized on the sphere in all dimensions.  The amount of null momentum that goes out through any local spherical cap is a fixed number, for each amplitude.

We'll characterize an excitation as localized near the geodesic in a large diamond, if it can fit into a small causal diamond, and evolve in the larger diamond, independently of the rest of the degrees of freedom in that diamond for the proper time along the geodesic.  Translated into quantum mechanics, this is a subset of fermion fields $\psi_{[i_1 \ldots i_9]} $, where $i_p$ runs between $1$ and $k$ .    In the empty diamond density matrix for the large diamond with size $ n \gg k$, these degrees of freedom are entangled with all of the others.  However, if we start in a state satisfying
\begin{equation}   \psi_{[1_1 \ldots i_8 J]} | S \rangle = 0 , \end{equation}  where $ i_p \leq k$ and $ J > k$ then the $U(n)$ currents have an upper triangular form and the interaction does not remove the constraint, to leading order in $1/N$. Since the entire evolution has a time scale of order $N L_P$, this means that, as $N \rightarrow\infty$ the number of constrained q-bits divided by $N$ cannot change.  The number of constrained q-bits divided by $N$ is thus an asymptotic conservation law. 
Starting at the largest diamond boundary in the past, the number of constrained q-bits is of order $N$ we see that as $N \rightarrow \infty$ the number of constrained q-bits divided by $N$ will be a continuous conserved quantum number, the energy in the rest frame of the particular geodesic we have chosen to focus on.  

Several interesting things can happen as we look at smaller and smaller diamonds on this geodesic.  The simplest is that the degrees of freedom associated with the localized object, and their associated constraints, are no longer included in the list of degrees of freedom associated with that diamond.   This would be the case if the object were actually at rest in a different Lorentz frame, either translated or boosted with respect to the original geodesic.  The conditions on overlapping causal diamonds tell us how we must relate the constraints appropriate to one geodesic to those on another.  For large diamonds, and with sufficient coarse graining it is easy to see how this works.  It is crucial that the background geometry gives us the framework for interpreting what is going on.  This is consistent with Jacobson's\cite{ted95} interpretation of that geometry as the hydrodynamic description of the quantum system under discussion.  Since the dynamics on the maximal area surface of each diamond is invariant under volume preserving maps, the constraints on a single geodesic don't tell us where the constrained q-bits lie on the surface.  It's only by reference to all of the geodesics surrounding it that one can establish where on each sphere of a nested family of diamonds the constrained q-bits lie.  The coarse grained solution to the consistency conditions is that they lie in $7$ dimensional annuli surrounding $8$ dimensional spherical caps.   Following the trail of those annuli through a nest of diamonds starting at the earliest time $-T$, one defines the incoming trajectory of a localized object through space-time.  This definition of a trajectory is more robust than that defined by perturbative quantum field theory.  We believe that the perturbative analog is a gravitational Wilson line, defining the center of the opening of a jet of soft gravitons surrounding a hard particle line, but this interpretation needs further investigation.   

\begin{figure}[h]
\begin{center}
\includegraphics[scale=0.5]{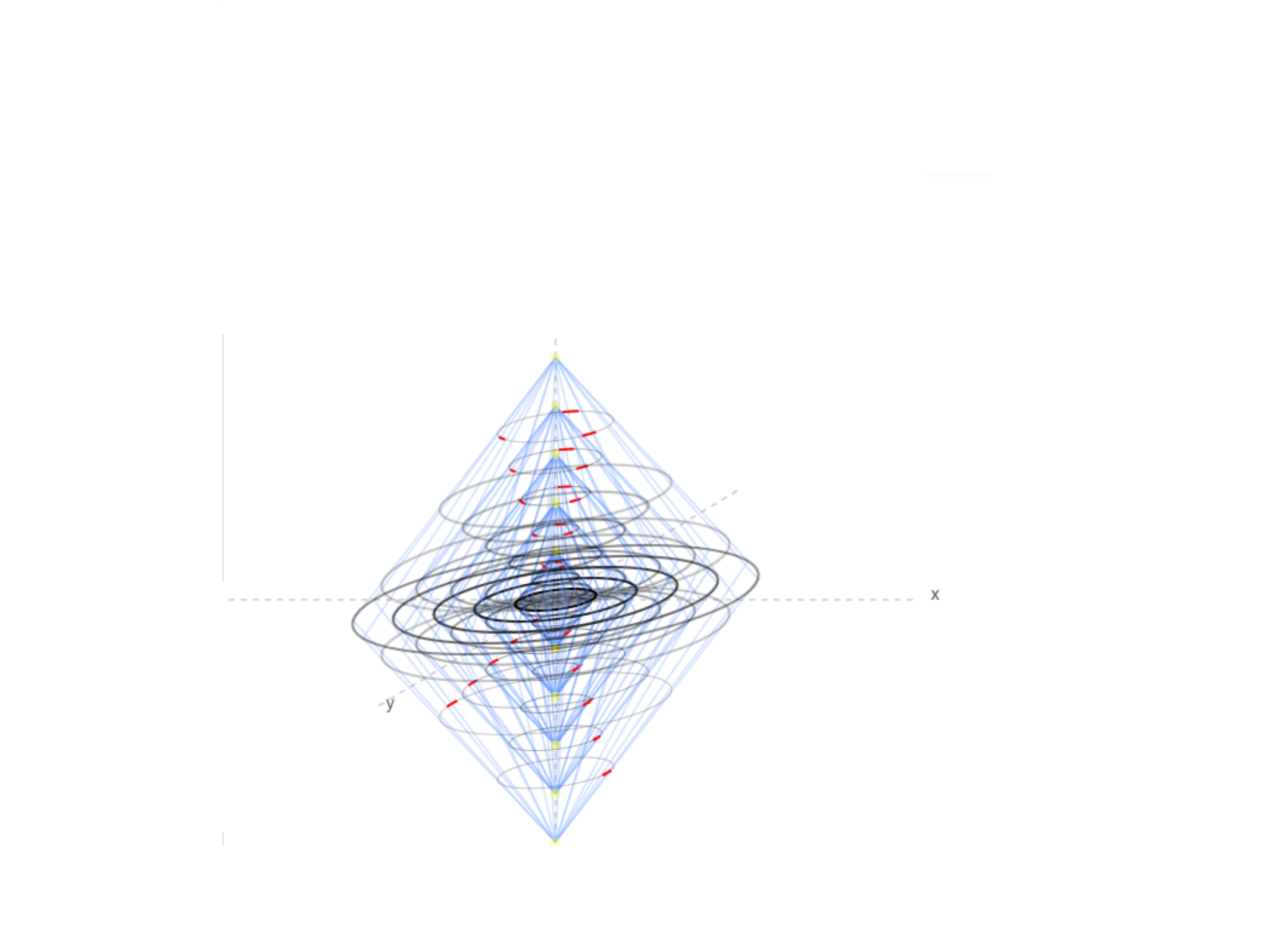}
\caption{2 +1 Dimensional Section of the Holographic Depiction of Jet Scattering: Red Arcs Are Q-bits Surrounded by "Frozen Q=bits" That Decouple Them From The Bulk of the Horizon Degrees of Freedom}
\label{z}
\end{center}
\end{figure}

The second interesting thing that can happen is that two or more sets of constraints can be imposed on the same diamond, in such a way that
\begin{equation} \sum k_j^8 n \sim n^9 .  \end{equation}  The diamond is then knocked far out of its empty diamond equilibrium state and the free Dirac Hamiltonian will quickly restore it to something close to equilibrium.   The nature of the resulting final state can be described differently, depending on the size of $n$.  If $n L_P$ is micro-scopically small the large number of constraints on the future boundaries of larger diamonds in the nest will still look like one or more trajectories of localized objects (Figure 2), but the number of outgoing objects does not have to be the same as the number of ingoing ones.  The causal diamond of size $n$ will look to a coarse grained observer like a local vertex where particle number changes.  These vertices can be connected together in large diamonds, to form the analog of time ordered Feynman diagrams. 

An example of this is shown in Figure 2 for $2 \rightarrow 3$ scattering.  The picture is misleading in several ways. Because of the scale, it's hard to make out the white spaces surrounding the red arcs denoting localized excitations as they enter or leave various causals diamonds.  Those spaces represent the frozen q-bits that allow the q-bits in the red arcs to evolve as independent systems for the time that they traverse the diamond.  The gray arcs represent the majority of the fluctuating q-bits in the diamond density matrix. Secondly, the topological properties of one dimensional circles make it appear that the angular positions of the localized objects are fixed.  In higher dimensions, the dynamics along a single geodesic is invariant under (fuzzy) volume preserving maps of the sphere.  It's only the consistency condition with the constraints on diamonds along other geodesics that fix the angular location.  If a constraint enters a diamond at a particular time, it must have come from an adjacent diamond the instant before that.  The position of that adjacent diamond tells us where to locate the center of the constrained q-bit annulus on the new diamond sphere.  

On the other hand, if $n \gg 1$ we have a large, high entropy equilibrium state.  If we take the equation $\sum k_i^8 \propto E$ literally, the equation for its formation is the equation for black hole formation in particle scattering.  Furthermore, the statistical probability for spontaneously finding this equilibrium state in a state where some set of 
$\psi_{[i_1 \ldots i_8 J]}$ with $i_j$ running up to $m$  is $\propto e^{- m^8 n}$, roughly what one expects from thermal Hawking decay.  Thus, its not unreasonable to take this as a model for black hole formation in particle scattering.  

Instead of defining our scattering matrix in terms of different states, we can build an extended operator algebra that incorporates the constraints.   Thus, if $P_{| s \rangle}$ is the projector on some particular scattering state we can define operators 
\begin{equation} P_{| s \rangle} \psi_{[i_1 \ldots i_9]} P_{| s \rangle} . \end{equation}  The scattering matrix is then a map between two isomorphic copies of the algebra of all of these operators, obtained by evolving the past copy with the time evolution operator along {\it any} geodesic from $- N L_P $ to $ N L_P$ and then taking the expectation value in the empty diamond state.  The $N \rightarrow\infty$ limit of the empty diamond state freezes out all but one momentum component of the $1 + 1$ dimensional fermion fields, since the geometry on the sphere at null infinity does not fluctuate.  

In order to take into account all possible scattering states we must first of all consider disjoint collections of $k_i$ indices, with $\sum k_i^8 \ll N^8$ , to define which fermion fields vanish on the initial state.   Then we must take into account that a given asymptotic injection of energy might not be retained in the causal diamond of the geodesic we have chosen to follow, because it is at rest in a frame related by a Poincare transformation.  This translates into a labelling of the initial projectors by an angle on $S^9$ and a magnitude on the null cone, where the magnitude is proportional to $k_i^8$ for the individual set of constraints.  In other words, we have a point on the lower half of the null cone.  There's a similar labelling for the projectors in the future algebra.  The limiting algebra is some kind of direct limit of these finite dimensional algebras.  The AGS algebra is a formal limit, in which the generators are taken to be operator valued half-measures on the null cone.  While we've not worked out the mathematical details, it seems clear that the limit should work.  The connection of our variables to the spinor bundle on $S^9$ tells us how to make combinations that are almost vanishing on selected subsets of the sphere for finite $N$.  Once we've done this for one annular region we can do it for any rotated one.  

\subsection{The $N \rightarrow\infty$ Limit}

The finite $N$ system we've described has a time evolution operator that maps between two copies of the same formal limiting operator algebra at the past and future boundaries of the maximal diamond $[- NL_P , N L_P]$.  This mapping $Q^+_a = S(Q^-_a )$ is obviously an algebra isomorphism because it is induced by a unitary operator $U_{in} (NL_P, - N L_P)$ for finite $N$, and we've {\it defined} a direct limit of the algebras that is invariant under the $11$ dimensional Poincare group.  We have not given a complete proof that the $S$ operation commutes with Lorentz transformations and conserves momentum, only partial arguments that time evolution is consistent when viewed from any geodesic and a strong argument that the energy in any geodesic rest frame is conserved.  

Apart from that the remaining question to ask is whether our limiting model is ``unitary".  It is far from clear that there is a representation of our limiting operator algebras in a separable Hilbert space.  On the other hand, for any finite $N$, our model is standard quantum mechanics in a finite dimensional Hilbert space and defines unitary transition amplitudes.  The literature on Algebraic Quantum Field Theory (AQFT) has developed a formalism to deal with situations similar to the one that faces us.  It involves the notion of a {\it locally normal state} on a net of operator algebras with a direct limit that may not have a representation in a separable Hilbert space.  Our situation is philosophically similar to those studied in AQFT but differs in mathematical detail.  In AQFT one wants to define the S-matrix for {\it e.g.} a local field theory with Nambu-Goldstone bosons, whose infinite volume limit does not have a separable Hilbert space.  One does this by defining a net of local operator algebras on large spheres, and {\it locally normal states} such that probability is conserved for any process where the incoming and outgoing energy-momentum flux is restricted to certain finite angular regions.  Then one can prove that the same is true in the limit of infinite size spheres.  The local algebras are Type $III_1$ von Neumann algebras and the proofs are intricate exercises in the theory of these notorious beasts.  Our setup does not have exact angular localization, because we impose a sharp angular momentum cutoff.  However, any real experiment has a finite angular resolution, so we can always choose $N$ large enough that our $Q_a (\Omega, p)$ operators are vanishing outside of the specified regions up to the experimental accuracy.  The bonus of our approach is that for finite volume, all operator algebras are finite dimensional and there are no intricate mathematical proofs to do.  Unitarity is obvious because we start from a finite dimensional quantum mechanical system with a well defined density matrix.  Our limiting AGS algebra is constructed explicitly in terms of a direct limit of finite sets of fermion creation and annihilation operators.   The $Q_a (\Omega, p)$ are built out of the $\psi_{[i_1 \ldots i_9]}$ by sandwiching them between projectors built from the algebra.  The analog of the locally normal state is the empty diamond state of Carlip and Solodukhin, except that we take the $1 + 1$ dimensional momentum cutoff down to a single momentum state as we take $N$ to infinity, in order to suppress fluctuations of the geometry of the $S^9$ at null infinity.  Since these momenta are discrete, this must be done in steps.  Alternatively we could introduce a classical scalar field that depended on the stretched horizon coordinates and coupled to the mass operator of the Dirac fields.  Dialing this coupling continuously to infinity with $N$ could decouple all but a single momentum mode.  

\section{Conclusions}

If the scattering operator in Minkowski space is a unitary operator in some separable Hilbert space then it has a spectral decomposition
\begin{equation} S = \int dP(E) S(E) . \end{equation} The dependence of the projection valued spectral measure on $E$ is completely determined by Poincare invariance.  The integral of $S(E)$ over any small microcanonical window centered at $E_0$ is a unitary operator. We have used robust semi-classical arguments in two separate kinematic regimes to argue that there is an order one probability in those regimes to produce a final state whose overlap with any Fock state with $\leq M$ particles goes to zero exponentially like a power of $E_0$ as $E_0 \rightarrow \infty$.  The first regime is one in which subgroups of initial particles have subenergies going to infinity and enter a region of size $K_i R_s (E^i_{sub}).$  The $K_i$ are chosen large enough so that no black holes are formed and we can trust soft theorems/eikonal methods to calculate the amplitudes.  The result, in space-time dimension $d \geq 5$ is well approximated by a normalizable coherent state whose typical graviton has wavelength of order $K_i R_s (E^i_{sub}).$ 

The second regime is one in which the majority of physicists in the field now believe particle scattering can produce multiple black holes.  One then assumes the final state in the decay of large black holes is well approximated, at the coarse grained statistical level, by classical general relativity supplemented with Hawking's thermal predictions.  The conclusions are roughly the same as in the regime above, though thermal states have even less overlaps with states of low particle number than eikonal coherent states.  

We claim that these arguments are not subject to correction by the UV completion of the Einstein-Hilbert action to a genuine theory of quantum gravity and have shown why neither perturbative superstring theory, BFSS matrix theory, or AdS/CFT contradict any of these assertions.  Finally we reviewed a well defined finite dimensional algebraic scattering theory, which has structural features designed to mimic known properties of quantum supergravity in 11 dimensional Minkowsi space.  We showed that the algebra of operators and constraints on the Hilbert space, which defined that model, formally converged to the AGS algebra of local SUSY generators on the null cone at infinity, and argued that unitarity of the finite dimensional ``S -matrix" was enough to guarantee the analog of algebraic unitarity in a {\it (fuzzy) locally normal state} on the limiting algebra.  The remaining issue for this model is Poincare covariance of the matrix elements.  We provided some intuitive arguments, but no definitive proof, of this crucial requirement. An important question related to this is the extent to which we should insist on microscopic validity of the consistency conditions for the Hilbert bundle.  We will leave the discussion of this somewhat philosophical question to an appendix.

\section{Appendix}

Arguments going back to at least the 1980s suggest that quantum field theory can only describe a miniscule fraction of the total number of states in a finite causal diamond. To maximize the number of states with a localized excitation at fixed diamond size we must create a black hole.  On the other hand, quantum measurement theory tells us that a detector must contain a robust semi-classical q-bit for every q-bit of the system it needs to measure, and every such decohered q-bit consists of a large number of microscopic q-bits.  Finally, the only way we know, mathematically, to construct models of such detectors is via (cut-off) QFTs.  Thus, it is in principle impossible to verify most of the details of a model of quantum gravity by local experiments.  Our mathematical models of quantum gravity in asymptotically flat and anti-de Sitter space have obscured this unfortunate fact.

We can ask whether those models themselves give us precise information about local causal diamonds.  Again, the answer appears to be no.  For the case of AdS space it is widely agreed that bringing the boundary of space in from infinity is equivalent to some sort of an ultraviolet cutoff.  The standard lore of quantum field theory tells us that the continuum model is insensitive to the details of the cutoff, so that unless we can find some preferred cutoff scheme there can be no universal answer to what is going on in a causal diamond of finite size.  Similarly, in asymptotically flat space it is well known even in massive quantum field theories that the S-matrix does not determine local Green functions.  In quantum gravity there are no local fields.  

These considerations allow one some leeway in imposing the consistency condition on overlap density matrices in the Hilbert bundle formulation of quantum gravity.  Suppose we found a mathematical model where the consistency condition was only satisfied asymptotically as the space-like separation between geodesics became large compared to $L_P$.  It would still give rise to the asymptotic symmetries of asymptotically flat and AdS models, and it's possible that no local experiment could ever distinguish such a model from one where the consistency conditions were satisfied exactly.  We might be more satisfied with the second model, but is there really a scientific justification for that satisfaction?  

\section{Acknowledgements}
The author would like to thank S. Shenker for criticism of his initial proposals for implementing fast scrambling in the Hilbert bundle formulation, and P. Draper, J. Maldacena and W.Fischler for critical reading of the manuscript.  This work was NOT funded by the Department of Energy but was supported in part by funds allocated to the NHETC by the State of New Jersey.   

\end{document}